\documentclass{appolb}
\usepackage{epsfig}

\begin{document}
\pagestyle{plain}
\title{COMPARISON OF $\delta$-AND GOGNY-TYPE PAIRING INTERACTIONS}
\author{A. Baran and K. Sieja
\address{Institute of Physics, University of M. Curie-Sk{\l}odowska, \\
ul. Radziszewskiego 10, 20-031~Lublin, Poland}}
\maketitle

\begin{abstract}
The matrix elements of the zero-range $\delta$-force and the finite range 
Gogny-type pairing force are compared. The strengths of the $\delta$-interaction 
for rare-earth nuclei are adjusted.
Pairing gaps resulting from different pairing interactions are compared 
to experimental ones. 
\end{abstract}


\bigskip
The pairing interaction plays a crucial role in the nuclear mean-field model calculations.
The most frequently used pairing type is the so-called monopole pairing interaction
({$g$=\rm {const}} model). More realistic residual interactions are those with state-dependent
matrix elements, e.g.  $\delta$- and Gogny forces.
Previously the Gogny-type pairing
force were studied within conventional Hartree-Fock-Bogoliubov (HFB) method \cite{Dobacz1} 
and in fully self-consistent Relativistic Hartree-Fock-Bogoliubov theory 
(RHB) \cite{Meng}. Here we investigate different kinds of pairing forces using the
Nilsson mean-field model and BCS approximation.
 
Contrary to the finite-range Gogny force,  the $\delta$-interaction is the zero-range force.
The zero-range nature of the pairing interaction tends to overestimate the coupling to
continuum states \cite{Krieger}. This defect doesn't occure when using the Gogny force which is,
however, cumbersome to handle; it involves sophisticated numerical techniques
while the $\delta$-force is relatively simple for numerical calculations. 
 The effect of the finite range for the $\delta$-force can be easily simulated 
by introducing an energy cutoff which plays the role
of an additional parameter similarly like in the case of the monopole pairing model
except that
the $\delta$-interaction is less sensitive to the cutoff.

These seemingly different pairing interactions should produce in fact similar
results- the coherence length of nucleonic Cooper pair is of the order
of size of a nucleus, therefore its structure should be insensitive to
the details of the interaction in the particle-particle channel \cite{Bulgac}.   
The aim of this paper is to show strong correlations and similarity
between matrix elements
of the pairing part of the Gogny force and of the $\delta$-pairing interaction. 
The pairing strengths of the $\delta$ force for rare-earth nuclei are determined from 
empirical odd-even mass
differences.
Additionally it is shown that the diagonal matrix elements of the $\delta$-pairing
interaction fulfill some 
dependences known from the monopole pairing strength analysis, therefore an alternative way
of adjusting $\delta$-pairing strengths is given.
In order to simplify the calculations we assume the Nilsson single particle potential
as a generator of single particle states and wave functions.

The finite range part of the Gogny force has the form \cite{Decharge}
\begin{equation}
\hat V_G(\vec r_{12})=\sum_{i=1,2}(W_i+B_i P_{\sigma}-H_i P_{\tau}-M_iP_{\sigma}P_{\tau})
e^{-\vec r_{12}^2/\mu_i^2},
\end{equation} 
where $P_\sigma$ and $P_\tau$ are the spin and isospin exchange operators respectively.
In all calculations the D1S parameters \cite{Berger} have been used and it is assumed
that only the isovector part of the interaction is active, {\it i.e.\/}, $P_\tau=1$ 
(proton-proton and neutron-neutron channel).
The zero-range $\delta$ interaction reads \cite{Krieger}
\begin{equation}
\hat V_{\delta}(\vec r_{12})=V_0({1-\vec\sigma_1\cdot\vec\sigma_2})/{4}\delta(\vec r_{12}).
\end{equation}
We limit further considerations to isovector pairing only (T=1).
Both interactions can be then written in a similar form 
\begin{equation}
\hat V(\vec r_{12})=V(|\vec r_1-\vec r_2|)(A+BP_\sigma),
\label{int}
\end{equation}
where the radial part is given by (see eg. \cite{Casten})
\begin{equation}
V(|\vec r_1-\vec r_2|)=\sum_{\lambda=0}^{\infty}V_\lambda(r_1-r_2)
\frac{4\pi}{2\lambda+1}\sum_\mu Y_{\lambda\mu}^*(r_1)Y_{\lambda\mu}(r_2).
\end{equation}
It is convenient to calculate the matrix elements of the interaction (\ref{int}) in 
$J$ basis \cite{Meng}:
\begin{equation}
V_{JMJ'M'}=\langle(ab)JM|\hat V(\vec r_{12})|(a'b')J'M'\rangle,
\label{eqv}
\end{equation}
where $a=(l_aj_a)$. 
In the case of the same kind of nucleons and for the pairing channel one obtains
\begin{eqnarray}
V_G=\langle(ab)00|\hat V_G(\vec r_{12})|(a'b')00\rangle=
\delta_{j_aj_b}\delta_{j_{a'}j_{b'}}\delta_{l_al_b}\delta_{l_{a'}l_{b'}}
\sum_{S=0,1; \lambda} (-1)^S(2S+1)\nonumber\\
(2l_a+1)(2l_b+1)
\sqrt{2j_{a'}+1}\sqrt{2j_{b'}+1}\,
V_\lambda(r,r')(A+B(2S-1))\nonumber\\
\left(
\begin{array}{ccc}l_a&\lambda&l_{a'}\\
              	     0& 0& 0	
\end{array}  
\right)^2
\left\{
\begin{array}{ccc}l_a&l_{a'}&\lambda\\
              	     l_{a'}& l_a& S	
\end{array}
\right\}
\left\{
\begin{array}{ccc}l_a&l_{a}& S\\
              	     \frac{1}{2}&\frac{1}{2}&j_a
	
\end{array}
\right\}
\left\{
\begin{array}{ccc}l_{a'}&l_{a'}&S\\
              	   \frac{1}{2}&\frac{1}{2}&j_{a'}
	
\end{array}
\right\}.
\label{gme}  
\end{eqnarray}
\begin{eqnarray}
V_{\delta}=\langle(ab)00|\hat V_G(\vec r_{12})|(a'b')00\rangle=
\delta_{j_aj_b}\delta_{j_{a'}j_{b'}}\delta_{l_al_b}\delta_{l_{a'}l_{b'}}
\delta(r-r')\frac{V_0}{r^2}\nonumber\\
\sqrt{2j_{a'}+1}\sqrt{2j_{b'}+1}
\sum_\lambda(-1)^{\lambda+j_a-j_{a'}}
\frac{2\lambda+1}{8\pi}
\left(
\begin{array}
{ccc}l_a&\lambda&l_{a'}\\
              	     0& 0& 0	
\end{array}  
\right)^2.
\end{eqnarray}
The allowed integer values for $\lambda$ in the sums are $|l_a-l_{a'}|\le\lambda\le l_a+l_{a'}$.

The simplest pairing model, {\it i.e.}, the monopole pairing model, assumes that all 
of the matrix elements
(\ref{eqv}) in the pairing hamiltonian are equal and state independent.

In order to simplify our studies the calculations were performed in Nilsson model \cite{Langanke}.
Formula (\ref{gme}) consists of two spin terms ($S=0$ and $S=1$). Numerical calculations
have shown that the $S=1$ term contributes less than $0.1\%$ of the total 
value and can be neglected in further calculations. 
\begin{figure}
\begin{center}
\includegraphics[scale=0.6]{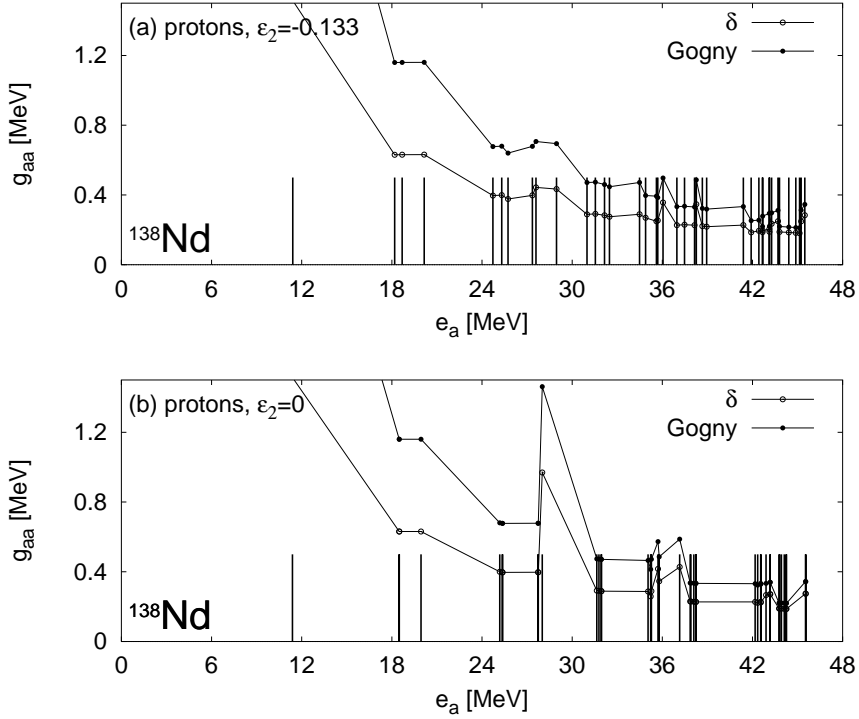}
\caption{Diagonal matrix elements $g_{aa}=(a\bar a|\hat V_{G|\delta}|\bar a a)$ v.s. single particle energies
for protons in $^{138}{\rm Nd}$.
Vertical lines represent the single particle spectrum.
The (a) part of the figure
corresponds to the equilibrium deformation while the part (b) to
the spherical case.}
\label{figel}
\end{center}
\end{figure}

\begin{figure}
\begin{center}
\includegraphics[scale=0.6]{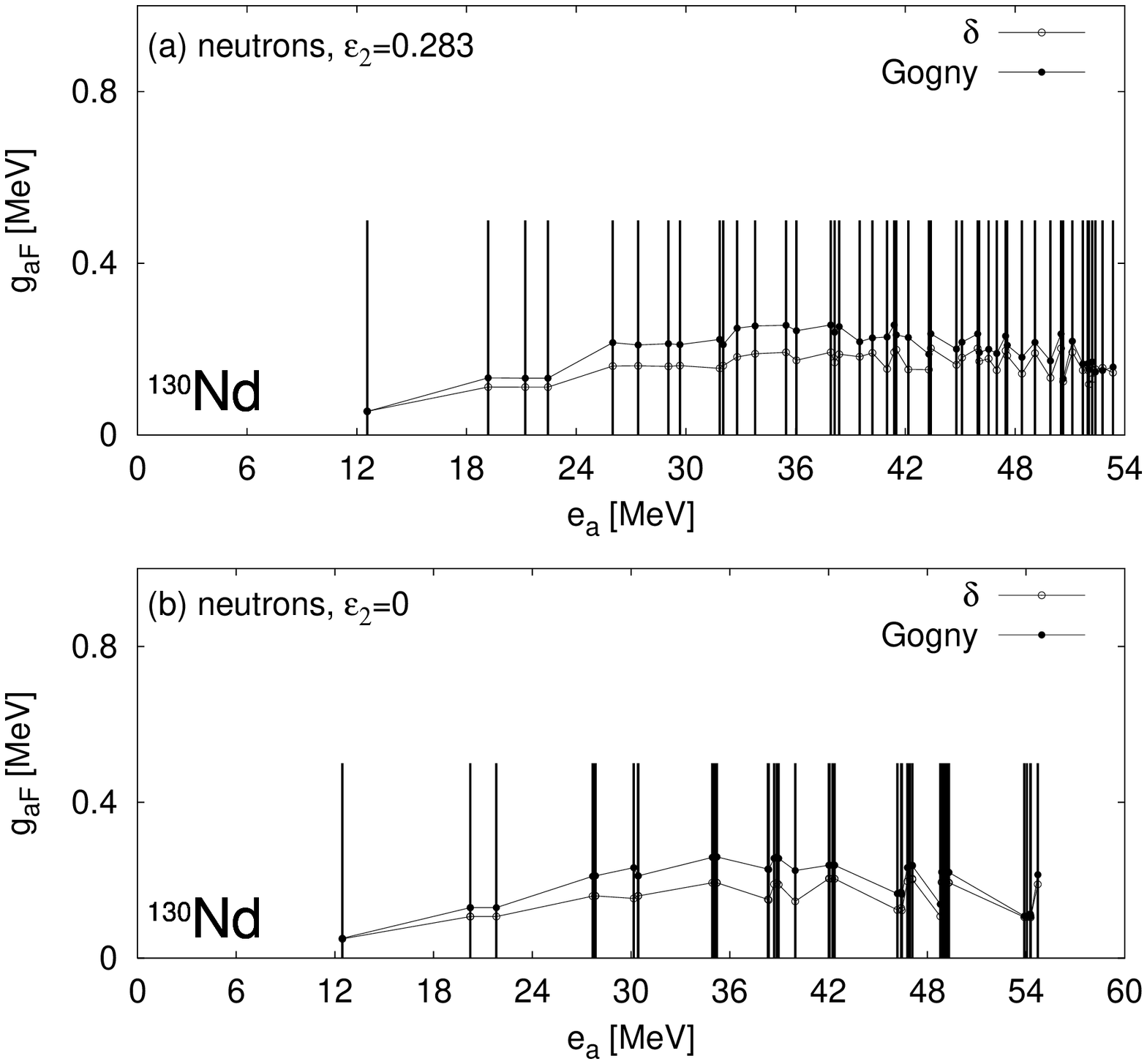}
\caption{Same as in Fig. \ref{figel} but for non-diagonal matrix elements
which couple to the Fermi level $F$ in the case of neutrons in $^{130}
{\rm Nd}$}.
\label{figeln}
\end{center}
\end{figure}
Figure \ref{figel} shows diagonal matrix elements 
($g_{aa}=(a\bar a|\hat V_{G|\delta}\bar a a)$) of
the Gogny force and of the $\delta$ interaction
for protons in $^{138}{\rm Nd}$, in cases of equilibrium and zero deformations. 
In Figure \ref{figeln} non-diagonal matrix elements are shown for neutrons 
in $^{130}{\rm Nd}$. 
We can obtain approximately the same values of matrix elements for both interactions considered here by a proper choice
of the $\delta$ force strength ($V_0$) which was not done in case of results shown
in Figures \ref{figel} and \ref{figeln} for the transparency. 
Strong correlations between diagonal and non-diagonal matrix elements of both interactions
are common for all considered nuclei, either for protons or for neutrons.

Usually the pairing strength is determined from odd-even mass differences
for separate nuclei.
In Ref. \cite{Boening} an explicit formula expressing the monopole pairing interaction strength
$g$ by the average gap parameter is derived and a simple dependence of $g$ on 
the neutron and proton numbers is found.
In the case of state-dependent pairing interactions
the diagonal matrix elements
follow the same particle number dependence as the monopole
pairing strength.

\begin{figure}[th]
\begin{center}
\includegraphics[scale=0.8]{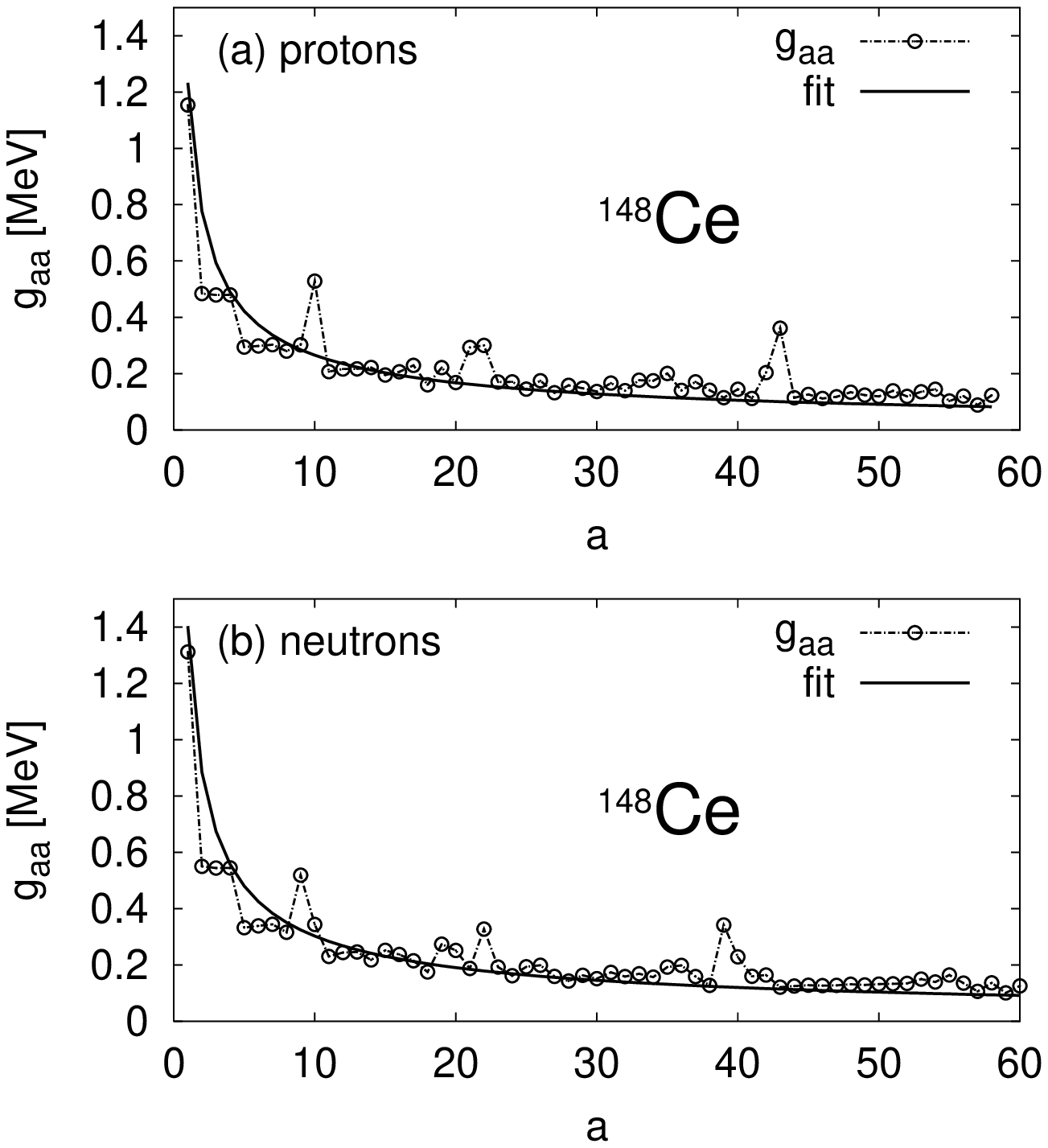}
\caption{Diagonal matrix elements $g_{aa}$ of the $\delta$ interaction as a function
of the level number $a$ for protons (a) and neutrons (b) in $^{148}{\rm Ce}$.
The solid lines represent the fit of the matrix elements according to Eq. \ref{fit}.}
\label{figdep}
\end{center}
\end{figure}

Figure \ref{figdep} shows the dependence of diagonal matrix elements $g_{aa}$ of the $\delta$-force on the particle number $2 n$ equal to the doubled number of levels $n$ 
(the possible number of particles occupying all of the levels up to the level $n$) for
the spectrum
$^{148}{\rm Ce}$. The dependence is of the type
\begin{equation}
g_{aa}(2n)={\rm const}(\hbar\omega_{0\ell})/{(2 n)}^{2/3}\,,
\label{fit}
\end{equation}
where $\hbar\omega_{0\ell}=41.0[1\pm(N-Z)/A]/A^{1/3}\, \rm{MeV}$.
The plus sign holds here for neutrons ($l=N$) and the minus for protons ($l=Z$).
The constant values found after the numerical fit of the $\delta$-pairing strength 
to the mass differences are 0.310 $\hbar\omega_{0N}$ for neutrons 
and 0.300 $\hbar\omega_{0Z}$ for protons. This behavior resembles the results
of Ref. \cite{Boening} where it was shown that the monopole constant $g_N=0.284\hbar\omega_{0N}$ and $g_Z=0.290\hbar\omega_{0Z}$.  This leads to 5$\%$ differences in the 
matrix elements. This may serve as an alternative way of adjusting an approximatre 
$\delta$-pairing strength and may be especially useful in case of nuclei far from the 
beta stability line, where no experimental masses are known.

Pairing gaps measure the lowest excitation energies of even-even nuclear systems and determine the
binding energy of the coupled pair of nucleons. There are a few different ways of
evaluating the pairing indicators from experimental nuclear masses. The three-point  
$\Delta^{(3)}$ indicator
was proved to be a proper estimate of the pairing gap parameter \cite{Dobacz}. It is defined as follows
\begin{equation}
\Delta^{(3)}(\mathcal N)=\frac{\pi_{\mathcal N}}{2}[B({\mathcal N}-1)
+B({\mathcal N}+1)-2B({\mathcal N})],
\label{3p}
\end{equation}
where $\pi_{\mathcal N}=(-1)^{\mathcal N}$ is the parity number and $B({\mathcal N})$ is the binding energy
of a system consisting of ${\mathcal N}$ particles.
Theoretical pairing gaps are calculated in BCS model. State dependent BCS equations
for the pairing gaps and the occupation probabilities are
\begin{equation}
\Delta_i=\frac{1}{2}\sum_jg_{ij}\frac{\Delta_j}{\sqrt{(e_j-\lambda)^2+\Delta_j^2}}\,,\,\,
\quad
v_i^2=\frac{1}{2}\left(1-\frac{e_i-\lambda}{\sqrt{(e_i-\lambda)^2+\Delta_i^2}}\right)\,.
\end{equation}
The set of BCS equations is solved numerically using a suitable iteration procedure.
The strengths $V_0$ of the $\delta$ force for protons and neutrons were adjusted for the 
pairing window containing $2\sqrt{15Z(N)}$ levels respectively. The values of $V_0$ were
determined from the condition of the equality of the experimental pairing indicator $\Delta^{(3)}$
and the lowest quasiparticle energy for each nucleus and then
the values of $V_0$ were averaged for the whole region.
The strengths of the $\delta$-pairing force for rare-earth nuclei obtained this way are \cite{ja}: 
$V_0^{protons}=240 {\rm MeV\cdot fm^3}$, 
$V_0^{neutrons}=230 {\rm MeV\cdot fm^3}$.
\begin{figure}
\begin{center}
\includegraphics[scale=1]{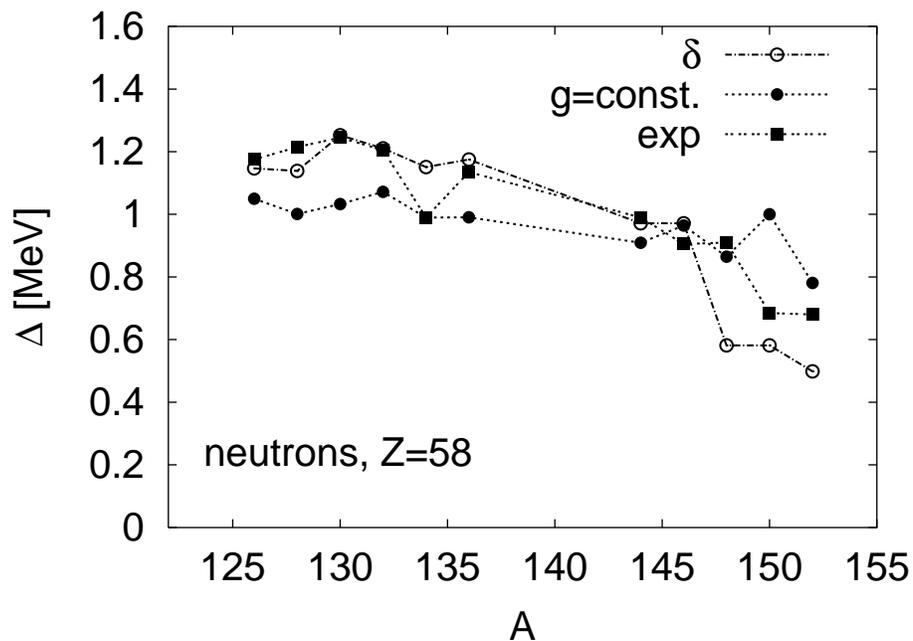}
\caption{Neutron pairing gaps calculated with the $\delta$-pairing interaction (opened circles)
and in the monopole pairing model (filled circles) in comparison to $\Delta^{(3)}$ pairing indicators
(squares).}
\label{gaps}
\end{center}
\end{figure}

Figure \ref{gaps} shows the pairing gaps for the $\delta$ interaction
(opened circles) and the monopole pairing interaction (filled circles)
 in comparison to experimental pairing gaps (Eq.\ref{3p})
for Cerium isotopes. The monopole pairing strengths used in the calculations are those of Ref. \cite{Boening}.
 The state dependent pairing results in theorethical pairing gaps which differ from $\Delta^{(3)}$ indicators 
on 10 $\%$ while
the monopole pairing gaps differ on 20 $\%$.
 
In conclusion, both the zero range $\delta$ force and the finite range part of the Gogny
 force have been investigated.
It was shown that after proper renormalization both forces give the same matrix elements
therefore they produce the same results. Thus the use of the $\delta$ force is justified and will
reduce numerical efforts in various studies.

The ${2 n}^{-2/3}$ dependence of the diagonal matrix elements
of the $\delta$-pairing force can be used in adjusting an approximnate 
$\delta$-pairing strengths.
s

\end{document}